\documentclass[aps,prx,reprint,superscriptaddress,nofootinbib]{revtex4-2}
\usepackage{blindtext}
\usepackage{lipsum}
\usepackage{graphics}
\usepackage{amsmath}
\usepackage{graphicx}
\usepackage{graphics}
\usepackage{amssymb}
\usepackage{verbatim}
\usepackage{physics}
\usepackage{float}
\usepackage{dsfont}
\usepackage[dvipsnames]{xcolor}
\usepackage{framed}
\definecolor{shadecolor}{RGB}{224,224,224}
\usepackage{amsthm}
\usepackage{amsfonts}
\usepackage{url}
\usepackage{chngcntr}
\counterwithout{equation}{section}

\makeatletter
\def\l@subsubsection#1#2{}
\makeatother

\usepackage[dvipsnames]{xcolor}
\usepackage[normalem]{ulem}
\usepackage{wasysym} 
\usepackage[export]{adjustbox}

\makeatletter
\DeclareFontFamily{OMX}{MnSymbolE}{}
\DeclareSymbolFont{MnLargeSymbols}{OMX}{MnSymbolE}{m}{n}
\SetSymbolFont{MnLargeSymbols}{bold}{OMX}{MnSymbolE}{b}{n}
\DeclareFontShape{OMX}{MnSymbolE}{m}{n}{
    <-6>  MnSymbolE5
   <6-7>  MnSymbolE6
   <7-8>  MnSymbolE7
   <8-9>  MnSymbolE8
   <9-10> MnSymbolE9
  <10-12> MnSymbolE10
  <12->   MnSymbolE12
}{}
\DeclareFontShape{OMX}{MnSymbolE}{b}{n}{
    <-6>  MnSymbolE-Bold5
   <6-7>  MnSymbolE-Bold6
   <7-8>  MnSymbolE-Bold7
   <8-9>  MnSymbolE-Bold8
   <9-10> MnSymbolE-Bold9
  <10-12> MnSymbolE-Bold10
  <12->   MnSymbolE-Bold12
}{}

\let\llangle\@undefined
\let\rrangle\@undefined
\DeclareMathDelimiter{\llangle}{\mathopen}%
                     {MnLargeSymbols}{'164}{MnLargeSymbols}{'164}
\DeclareMathDelimiter{\rrangle}{\mathclose}%
                     {MnLargeSymbols}{'171}{MnLargeSymbols}{'171}
\makeatother

\usepackage{tikz}
\usepackage{tikz-cd}
\usetikzlibrary{arrows}
\usetikzlibrary{snakes}
\usetikzlibrary{intersections}
\usetikzlibrary{shapes.geometric}
\usetikzlibrary{decorations.pathmorphing, patterns,shapes}
\usetikzlibrary{decorations.markings}
\usetikzlibrary{calc}

\tikzset{
	partial ellipse/.style args={#1:#2:#3}{
		insert path={+ (#1:#3) arc (#1:#2:#3)}
	}
}

\tikzset{
	mid arrow/.style={postaction={decorate,decoration={
				markings,
				mark=at position .575 with {\arrow[#1]{stealth}}
	}}},
	near arrow/.style={postaction={decorate,decoration={
				markings,
				mark=at position .275 with {\arrow[#1]{stealth}}
	}}},
	far arrow/.style={postaction={decorate,decoration={
				markings,
				mark=at position .800 with {\arrow[#1]{stealth}}
	}}},
}

\pgfdeclarepatternformonly{south west lines}{\pgfqpoint{-0pt}{-0pt}}{\pgfqpoint{3pt}{3pt}}{\pgfqpoint{3pt}{3pt}}{
	\pgfsetlinewidth{0.4pt}
	\pgfpathmoveto{\pgfqpoint{0pt}{0pt}}
	\pgfpathlineto{\pgfqpoint{3pt}{3pt}}
	\pgfpathmoveto{\pgfqpoint{2.8pt}{-.2pt}}
	\pgfpathlineto{\pgfqpoint{3.2pt}{.2pt}}
	\pgfpathmoveto{\pgfqpoint{-.2pt}{2.8pt}}
	\pgfpathlineto{\pgfqpoint{.2pt}{3.2pt}}
	\pgfusepath{stroke}}

\definecolor{orange(ryb)}{HTML}{FFA500}
\definecolor{lightorange(ryb)}{HTML}{FFB300}
\definecolor{dodgerblue}{HTML}{1E90FF}
\definecolor{lightdodgerblue}{HTML}{4dbff7}
\definecolor{crimson}{HTML}{FF4C4C}
\definecolor{pinkerton}{HTML}{EC368D}
\definecolor{forest}{HTML}{6DD189}
\definecolor{lightishgray}{HTML}{DFDFDF}
\definecolor{error-red}{HTML}{EFB2B6}

\usepackage[colorlinks=true,citecolor=dodgerblue,linkcolor=dodgerblue,urlcolor=dodgerblue,pdftitle={Mielke}]{hyperref}

\def \beq {\begin{equation}}
\def \eeq {\end{equation}}
\def \beqa {\begin{eqnarray}}
\def \eeqa {\end{eqnarray}}
\def \bseq {\begin{subequations}}
\def \eseq {\end{subequations}}

\newcommand \down {\downarrow}

\begin{document}

\title{Spin Polarons in Flat Band Ferromagnets}

\author{Saranesh Prembabu}
\affiliation{Department of Physics, Harvard University, Cambridge, MA 02138, USA}
\author{Rahul Sahay}
\affiliation{Department of Physics, Harvard University, Cambridge, MA 02138, USA}
\author{Stefan Divic}
\affiliation{Department of Physics and Astronomy, University of Pennsylvania, PA 19104, USA}
\author{Ashvin Vishwanath}
\affiliation{Department of Physics, Harvard University, Cambridge, MA 02138, USA}

\begin{abstract}
Spin polarons are bound states of electrons and spin-flips that form above spin polarized electronic insulators.
These bound states conventionally form in one of two settings: in frustrated lattices with dispersive bands---where the motion of an electron preferences binding a nearby spin-flip---or in topological flat bands---where the Chern number enforces an effective dipolar interaction between electrons and spin flips.
In this work, we report the formation of a spin polaron in a context that doesn't fall cleanly into either of these paradigms.
In particular, we study the one-dimensional Mielke-Tasaki chain, a paradigmatic model of flat band ferromagnetism, which has an exact  ferromagnetic ground state, trivial band topology, and quenched kinetic energy in its lowest band.
Despite these features, our density matrix renormalization group simulations reveal the presence of spin polarons upon electron doping this model.
More surprisingly, combining these numerics with analytic calculations, we show that polaron binding occurs when the interaction-induced kinetic energy of the model is zero---contrary to intuition from kinetic magnetism---and the glue binding the electrons and spin-flips arises from weak mixing with the model's dispersive band---contrary to what occurs in topological flat bands.
Our results open the doors to exploring how the quantum geometry of flat bands drives the formation of exotic charge carriers.

\end{abstract}

\maketitle

\textbf{Introduction.} The low-energy charge carriers of a correlated material need not resemble its constituent electrons.
For instance, in some circumstances, these charge carriers can instead be composite objects built from electrons bound to other emergent quasi-particle excitations present in the system.
As a key example, in systems with spin-polarized ground states, these composites can be \textit{spin polarons}---i.e., electrons or holes bound to spin flips atop the polarized background.

Such excitations have recently attracted considerable attention.
On the experimental front, these polarons and their multi-magnon generalizations (``skyrmions'')~\cite{skyrme1962unified, belavin1975metastable, sondhi1993skyrmions, macdonald1996skyrmions, lee1990boson, Khalaf_2022, Schindler_2022} have been observed in a wide range of experimental platforms, ranging from moir\'e and cold atom Fermi-Hubbard simulators~\cite{tang2020simulation, Ciorciaro_2023, tao2023observationspinpolaronsfrustrated, Lebrat_2024, Ji_2021, qiao2025kineticallyinducedboundstatesfrustrated, prichard2025observationmagnonpolaronsfermihubbardmodel, Prichard_2024} to topological settings with strong magnetic fields~\cite{liu2022visualizing,girvin1999quantumhalleffectnovel}.
Moreover, recent theoretical work has found that spin polarons in a material can pair upon doping leading to superconductivity from purely repulsive interactions~\cite{Chatterjee_2022, Grover_2008, wang2025spinpolaronmediatedsuperconductivitydoped, wang2025intertwinedordersquantumcriticality}.
As a consequence, such excitations can be viewed as resources for the realization of superconducting ground states.

These experimental observations and theoretical prospects naturally motivate investigating the different mechanisms and settings in which such spin polarons can arise.
Most known routes to polaron formation relate to one of two mechanisms.
The first is kinetic magnetism, whereby a charge carrier on a frustrated lattice (e.g., the triangular lattice) lowers its kinetic energy by being dressed with spin flips~\cite{nagaoka1966ferromagnetism, Schrieffer_1988, haerter2005kinetic, sposetti2014classical, lisandrini2017evolution, Zhang_2018, Davydova_2023, zhang2023pseudogap,Morera_2024, Morera_Navarro_2024, morera2024itinerantmagnetismmagneticpolarons, Samajdar_2024}.
The second mechanism arises in topological flat bands, wherein the Chern number of a band enforces an effective dipolar interaction between electrons and spin flips, binding them into polarons~\cite{sondhi1993skyrmions, Khalaf_2022, macdonald1996skyrmions, Schindler_2022, girvin1999quantumhalleffectnovel}.

\begin{figure}[!t]
    \centering
    \includegraphics[width = 247 pt]{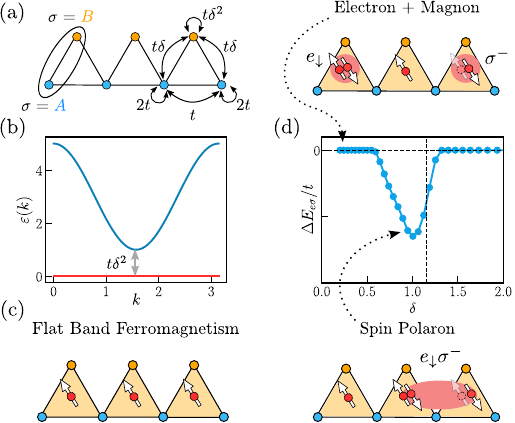}
    \caption{\textbf{Spin Polaron Formation in a Flat Band Ferromagnet.} \textbf{(a)} We explore spin polaron formation in the 1D Mielke-Tasaki chain---a model of repulsively interacting electrons on the sawtooth lattice whose hoppings are set by a dimensionless parameter $\delta$.
    \textbf{(b)} At all $\delta$, this model has an exact flat band (red) as well as a dispersive band (blue) separated by a gap $t\delta^2$.
    \textbf{(c)} Also, the ground state at half-filling of the flat band is an $SU(2)$ symmetry-breaking ferromagnet for \textit{any} repulsive on-site interaction $U>0$. 
    \textbf{(d)} We study the nature of low-energy charge carriers doped above the flat-band ferromagnet by numerically computing the binding energy $\Delta E_{e \sigma}$ [defined in Eq.~\eqref{eq:binding}] between doped electron and spin flip (magnon) excitations of the ferromagnet. 
    In some regimes of $\delta$ and the interaction strength $U$, this binding energy is zero and consequently electrons are the lowest energy charge carriers and remain decoupled from any added spin flips in the system (shown on top).
    Importantly, in other regimes, the binding energy is negative and consequently spin polarons (shown on bottom) form. 
    }
    \label{fig:overview}
\end{figure}

In this work, we demonstrate the formation of a spin polaron in a context that \textit{does not} fall cleanly into either the framework of kinetic magnetism or band topology.
Namely, we find that a spin polaron arises due to the presence of repulsive interactions in the one-dimensional Mielke-Tasaki (MT) chain~\cite{Mielke_1992, tasaki1998nagaoka, AMielke_1991, Mielke1991, Mielke1993443, Tasaki_1994, Mielke_1993, mielke1996note, mielke1999stability, Tasaki_1992, tasaki1998hubbard}---a paradigmatic model of flat-band spontaneous ferromagnetism which, crucially, has trivial band topology and a \textit{quenched} bare kinetic energy of charge carriers [c.f. Fig.~\ref{fig:overview}(a-c)]. 
In what follows, we first use the density matrix renormalization group (DMRG) to establish that, at half-filling of the model's lowest (flat) band, a spin polaron appears as a bound state above the ferromagnetic ground state.
Subsequently, through a combination of numerics and analytics, we explore the energetics of this polaron formation; we provide explanations for where it forms in the MT chain's parameter space and demonstrate that the glue binding the electron and spin flip is a consequence of a weak interaction-induced hybridization with the MT chain's upper (dispersive) band. 
Surprisingly, we find that polarons are formed when their renormalized (interaction-induced) dispersion is quenched.
Our work opens the doors to an understanding of polaron formation outside settings with either topological bands or kinetic magnetism.

\textbf{The Mielke-Tasaki Model.} The MT model~\cite{Tasaki_1992} is a one-dimensional system of interacting fermions that live on the bipartite sawtooth lattice shown in Fig.~\ref{fig:overview}(a).
If $\psi^{\dagger}_{x, \sigma, s}$ labels the creation operator of a fermion with spin $s$ living at unit cell $x$ and sublattice $\sigma \in \{A, B\}$, then the model can be succinctly expressed as:
\begin{equation} \label{eq-MTHam}
    H = t \sum_{x, s} \Psi^{\dagger}_{x, s} \Psi_{x, s} + U \sum_{x, \sigma} n_{x, \sigma, \uparrow} n_{x, \sigma, \downarrow}, 
\end{equation}
where $n_{x, \sigma, s} = \psi_{x, \sigma, s}^{\dagger} \psi_{x, \sigma, s}$ is the electron density, $U, t>0$ are the on-site interaction strength and hopping coefficient respectively, and $\Psi_{x,s} = \left( \psi_{x,A,s} + \psi_{x + 1, A,s} + \delta \psi_{x, B,s} \right)$ is a linear combination of the fermion operators that depends on a dimensionless parameter $\delta > 0$.
The single-particle term in Eq.~\eqref{eq-MTHam} concisely encodes the nearest-neighbor hopping and on-site energies shown in Fig.~\ref{fig:overview}(a), wherein electrons feel a potential of $2t$ on the $A$ sublattice and $t\delta^2$ on the $B$ sublattice, while the inter- and intra-sublattice hopping amplitudes are $t\delta$ and $t$, respectively. The dimensionless ratio $\delta$ determines the sublattice distributions of flat band states and, in what follows, will be a tuning knob into qualitatively different regimes of the model.

In our investigation, we choose the Mielke-Tasaki model for two reasons.
First, at the single particle level, the model consists of two topologically trivial bands per spin species in the Brillouin zone $[0,\pi)$, the lower of which is exactly flat and separated from the dispersive upper band by a gap $t \delta^2$ [see Fig.~\ref{fig:overview}(b)] \cite{Tasaki_1992}.
Second, the single-particle term in Eq.~\eqref{eq-MTHam} is manifestly positive semi-definite.
Together with $U\ge 0$, this implies that any fully spin-polarized Slater determinant filling the flat band is a ground state [c.f. Fig.~\ref{fig:overview}(c) for a schematic].
Crucially, such states \textit{spontaneously} break the $SU(2)$ spin rotation symmetry of the Hamiltonian.
Rigorous results further show that, for any arbitrarily small repulsive $U>0$ and quarter filling on a periodic chain of $2N$ sites, these saturated ferromagnetic states are unique ground states up to its $SU(2)$ multiplet of dimension $2S+1=N+1$ with total spin $S=N/2$, where $N$ is both the number of unit cells in the periodic chain and the number of electrons at half filling of the lowest band~\cite{Mielke_1992, tasaki1998nagaoka, AMielke_1991, Mielke1991, Mielke1993443, Tasaki_1994, Mielke_1993, mielke1996note, mielke1999stability, Tasaki_1992, tasaki1998hubbard}.
As an important added remark, the ferromagnetic ground state of the MT chain has a spin stiffness that scales linearly with the interaction strength~\cite{tasaki1998nagaoka}, in spite of being a flat band, a consequence of its quantum geometry~\cite{girvin1999quantumhalleffectnovel, QGStiffness}.

Prior to launching into our numerical investigations into polaron formation, we highlight key features of the model that are similar to and distinct from other settings in which polarons arise.
In particular,  the triangular structure and positive hopping coefficient $t$ of the Mielke-Tasaki model is reminiscent of the Fermi-Hubbard model on the triangular lattice, where spin polarons have previously been observed~\cite{nagaoka1966ferromagnetism, Schrieffer_1988, haerter2005kinetic, sposetti2014classical, lisandrini2017evolution, Zhang_2018, Davydova_2023, zhang2023pseudogap,Morera_2024, Morera_Navarro_2024, morera2024itinerantmagnetismmagneticpolarons, Samajdar_2024}.
There, polarons arise in the strongly interacting limit and in the presence of a weak spin-polarizing Zeeman field as a consequence of the frustrated motion of electrons.
In contrast, in the Mielke-Tasaki model, the spin-polarized ground state arises spontaneously and, more importantly, since the lowest band of the model is flat, the motion of electrons is quenched. 
While flat band spin polarons are known to arise in the context of quantum Hall and Chern ferromagnets, the origin of these is intimately tied to the Chern number of these bands as described in Ref.~\cite{Khalaf_2022}.
This is in contrast to the case of the MT model, whose bands are one-dimensional and topologically trivial.
In what follows, we demonstrate that the MT model nevertheless exhibits a weakly bound spin polaron excitation.

\begin{figure}
    \centering
    \includegraphics[width=247pt]{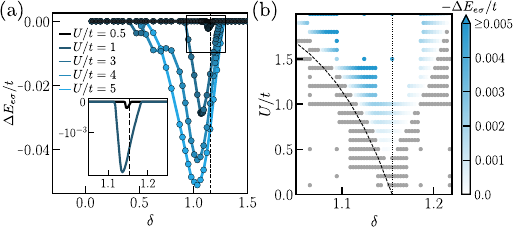}
    \caption{\textbf{Numerical Evidence of Polaron Formation}. By simulating the Mielke-Tasaki chain using DMRG, we find a window of hopping ratios $\delta$ and a threshhold interaction strength $U$ above which the model hosts a spin polaron bound state.
    \textbf{(a)} In particular, we first plot the binding energy of the polaron $\Delta E_{e\sigma}$ [see Eq.~\eqref{eq:binding}] as a function of $\delta$ for different values of $U$ above the observed threshold value of $U_c/t \approx 0.43$.
    We find a negative $\Delta E_{e\sigma}$ in a window of $\delta$, indicating binding of electrons and spin-flips. 
    We note that at the weakest interaction strength, where the physics of the flat band plays the largest role, a polaron first forms around $\delta_c \approx 1.155$ (see inset and vertical dashed line).
    \textbf{(b)} We summarize our numerical investigations by plotting the negative of the polaron binding energy, shown in blue (gray) at parameters with (without) polaron formation.
    We provide a theoretical estimate of the threshold $U$ for $\delta < \delta_c$ (dashed curve) based on considerations of the interaction-induced dispersion of the electron, which we explain in detail in the section on polaron energetics.
    }
    \label{fig:numerical-evidence}
\end{figure}

\textbf{Numerical Evidence of a Spin Polaron}---To demonstrate the existence of a spin polaron, we start by recalling that, given the $U(1)_{S_z} \subset SU(2)$ spin rotation symmetry and $U(1)_N$ charge conservation symmetry of the MT model, we can label each of its eigenstates by the total number of spin up and down electrons, $N_{\uparrow}$ and $N_{\downarrow}$.
Consequently, working in the sector $(N_{\uparrow}, N_{\downarrow}) = (N, 0)$, where  $N$ is the total number of unit cells in the chain, the ground state is given by fully filling the lowest (flat) spin-up band. We choose periodic boundary conditions to eliminate edge effects.

To check for the existence of the spin polaron, one must compare the lowest energy excitation in the sector containing both an additional flat-band (spin-down) electron and a spin flip [i.e., $(N_{\uparrow}, N_{\downarrow}) = (N -1, 2)$] to the lowest energy excitations in the sectors independently containing the electron and the spin flip [i.e., $(N_{\uparrow}, N_{\downarrow}) = (N, 1)$ and $(N_{\uparrow}, N_{\downarrow}) = (N-1, 1)$].
Denoting the excitation energies of these sectors relative to the spin-polarized ground state as $E_{e\sigma}$, $E_{e}$, and $E_{\sigma}$ respectively, it thus suffices to check that there is finite binding energy:
\begin{equation} \label{eq:binding}
     \Delta E_{e \sigma} \equiv E_{e\sigma} - E_{e} - E_{\sigma} < 0.
\end{equation}
A key observation is that, since the MT chain \textit{spontaneously} breaks the $SU(2)$ spin rotation symmetry, the magnon excitation---obtained via the action of the spin-lowering operator---is a gapless quadratically-dispersing goldstone mode~\cite{PhysRevB.110.195140}.
Consequently, $E_{\sigma} = 0$ and hence only $E_{e \sigma}$ and $E_{e}$ must be compared. As an aside, we do not see analogous polaron binding energy for holes, because the energy of a decoupled hole already  is zero.

To do so, we perform finite DMRG on the MT chain, conserving both electronic charge and $S^z$ spin.
We compute the ground state energy in the sector of the spin polaron $(N_{\uparrow}, N_{\downarrow}) = (N - 1, 2)$ and the sector of the electron $(N_{\uparrow}, N_{\downarrow}) = (N, 1)$ to determine a putative polaron binding energy as a function of the parameter $\delta$ [in Eq.~\eqref{eq-MTHam}] and interaction strength $0.1\lesssim U/t \lesssim 5$ (a larger range is explored in \cite{SM}).
Specifically, we focus on the small $U$ regime as we are primarily interested in physics dominated by the flat band.
Our numerics are performed at a system size of $2N = 40$ lattice sites and a bond dimension of $\chi = 256$, which we found to be sufficient for convergence. Doubling the length or bond dimension was found to not affect the measured energies appreciably beyond our numerical error threshold.
The binding energy is reported as a function of $\delta$ and different interaction strengths $U$ in Fig.~\ref{fig:numerical-evidence}(a) and we summarize our numerical observations in Fig.~\ref{fig:numerical-evidence}(b), where lack of polaron formation is indicated in gray and the strength of the binding when polarons are formed is indicated with a blue gradient (with a numerical threshold $\Delta E_{e\sigma}/t \gtrsim 10^{-8}$).

We find that spin polarons form in a particular window of hopping ratios $\delta$ and only above a certain threshold repulsive strength $U$ [Fig.~\ref{fig:numerical-evidence}(b)].
In particular, as we increase $U$ from zero, the onset of spin polaron formation is first seen at $\delta_c \approx 1.15$ once $U/t$ exceeds a threshhold value of $U_c/t \approx 0.43$ [drawn with a dotted line in Fig.~\ref{fig:numerical-evidence}(a,b); see inset of Fig.~\ref{fig:numerical-evidence}(a) for the onset].
The onset occurs very sharply and sensitively with respect to $\delta$, such that increasing or decreasing $\delta$ by a few percent can double the threshold $U$ needed for spin polaron formation.
Moreover, the binding energy reaches a maximum of roughly $-0.05t$ around $U/t \approx 6$, then decreases and eventually vanishes for stronger interactions \cite{SM}.
If $\delta$ deviates too far above or below $\delta_c$ (namely, $\delta\lesssim 0.3$ or $\delta \gtrsim 1.8)$ a polaron does not form at any $U$.

\textbf{Energetics of Polaron Formation.} The observation of a spin polaron in this context is surprising as it arises in a context that doesn't cleanly fit into either the framework of kinetic magnetism or topological bands.
Moreover, several features of our numerical observations warrant explanation---notably, (1) What sets the critical value of $\delta$ observed? (2) What sets the scale of the threshold interaction strength $U$ at which the polaron forms at each $\delta$? (3) What is the glue that binds electrons and magnons into polarons?
We now turn to addressing these questions.

To do so, let us start by making a general comment on the energetic requirements for polaron formation.
In particular, note that, in the absence of an effective attractive interaction between electrons and spin flips, it is energetically preferable for each excitation to localize in momentum space at their respective band minima.
Consequently, for polarons to form, a sufficiently large attractive interaction between electrons and spin-flips must enable them to overcome the energetic cost of delocalizing in momentum space to form a bound state.
In what follows, we shed light on the above questions by 
analyzing the energetics of the electrons and magnons of the Mielke-Tasaki model as a function of $\delta$ and $U$.

\begin{figure}
    \centering
    \includegraphics[width=247pt]{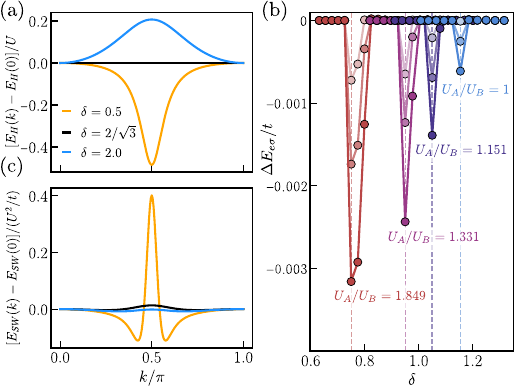}
    \caption{\textbf{Polaron Formation and Interaction-Induced Dispersion}. In contrast to intuition from kinetic magnetism \cite{Schrieffer_1988, haerter2005kinetic, sposetti2014classical, lisandrini2017evolution, Zhang_2018, Davydova_2023, Morera_2024, Morera_Navarro_2024, morera2024itinerantmagnetismmagneticpolarons}, in the Mielke-Tasaki model, polarons form when the interaction-induced dispersion of electrons is minimized. 
    \textbf{(a)} To see this, we first plot the interaction-induced (Hartree) dispersion of doped electrons that arises due to their interaction with the density of the background ferromagnet, $[E_H(k)-E_H(0)]/U$.  
    This dispersion is \textit{perfectly flat} for $\delta < 2/\sqrt{3} \approx 1.155$, precisely where we see the polaron arise at the weakest interaction strengths.
    In contrast, polarons form at much larger interaction strength when $\delta \neq \delta_c$ where the dispersion is sizable.
    \textbf{(b)} To solidify the relationship between polaron formation and a flat Hartree dispersion, we show that one can tune the location $\delta$ where the dispersion vanishes using a staggered interaction [discussion below Eq.~\eqref{eq:Hartreedensity}].
    By plotting the polaron binding energy as a function of $\delta$ for different staggering ratios $U_A/U_B$ and a weak average interaction $(U_A + U_B)/2 \approx \{0.45,0.60,0.75\}$ (lightest to darkest), we find that the onset of polaron formation tracks the location that the dispersion vanishes $\delta \approx 1.155, 1.05,0.95,0.75$ (vertical dashed lines). 
    \textbf{(c)} Furthermore, when $\delta < \delta_c$, we find that the second-order (Schrieffer–Wolff) correction to the single electron dispersion $E_{\mathrm{SW}}(k)$ can approximately cancel the Hartree dispersion at a sufficient interaction strength due to its peaked shape.
    This provides a predictive heuristic for the onset of polaron formation in $U$ for $\delta < \delta_c$, shown in Fig.~\ref{fig:numerical-evidence}(b).
    }
    \label{fig:3}
\end{figure}

\textit{Location in $\delta$: Overcoming Single-Electron Energetics}---To understand why the polaron first forms around $\delta_c \approx 1.155$, it is necessary to understand the energetics of a single electron doped atop the ferromagnetic ground state.
In particular, in the weakly interacting regime, the physics of the system largely takes place within the model's flat band.
Here, while the single-particle energy of an added electrons is zero, the interaction between an added electron and non-uniformities of the background density of the ferromagnet can give it a non-trivial dispersion.
This can be made precise by projecting the MT chain into its flat band, which yields: 
\begin{equation}
    \mathcal{P}H\mathcal{P} = U \sum_q \rho_{q\uparrow}\,\rho_{-q\downarrow}, 
\ \  
\rho_q = \sum_k  \lambda_q(k)c^\dagger_k c_{k+q},
\label{eq:flatbandformalism}
\end{equation}
where $\mathcal{P}$ is the many-body projector onto the flat band, the form factor $\lambda_q(k) =\langle u_k|u_{k+q}\rangle$ is the overlap of Bloch wave vectors, and $c_k$ annihilates a flat-band electron at momentum $k$.
Evaluating this Hamiltonian on the space of single electrons atop the ferromagnet yields: 
\begin{equation} \label{eq:Hartreedensity}
    \mathcal{P}H \mathcal{P} \ket{k} = E_H(k) \ket{k}, \ \  E_H(k) = U \big\langle u_k \big| 
\begin{pmatrix}
n_A & 0 \\[2pt] 0 & n_B
\end{pmatrix}
\big| u_k \big\rangle,
\end{equation}
where $\ket{k}$
is the state corresponding to a single added spin-down electron at momentum $k$ above the spin-up ferromagnetic ground state, $n_A$ and $n_B$ (with $n_A+n_B=1$) are the fractions of the ferromagnetic background on sublattices $A$ and $B$, respectively. Moreover, $E_H(k)$ is the ``Hartree dispersion'' and is depicted for different $\delta$ in Fig.~\ref{fig:3}(a)\footnote{We remark that the Fock contribution to the dispersion is zero in this context because a doped spin down elecron can't exchange with the oppositely polarized background.}
The above makes manifest the fact that it is the non-uniformity of the background  electron density on the two sublattices that leads to a dispersion.
One finds from the Bloch wavefunctions that $n_A=\delta/\sqrt{\delta^2+4}$ and $n_B=1 - n_A$, consistent with the intuitive expectation that $\delta$ polarizes electrons away from the $B$ sublattice.
Crucially, $n_A = n_B = 1/2$ exactly at $\delta_c = 2/\sqrt{3} \approx 1.155$, precisely the $\delta$ value where the polaron forms (within $\sim 0.005$).
This leads to a clear physical explanation for where the polaron forms at $\delta$: at $\delta_c$, the interaction-induced dispersion in the flat band is flat and consequently, any putative attractive interaction between electrons and magnons can freely delocalize the electron across momentum space.

This understanding yields an intriguing and concrete prediction.
To set the stage 
, note that for \textit{any} $\delta$ we can modify the Hamiltonian interactions so that the Hartree dispersion becomes flat at that specific $\delta$.
This can be achieved straightforwardly by introducing a staggered repulsion strength—assigning different interaction parameters $U_A$ and $U_B$ to the $A$ and $B$ sublattices, respectively.\footnote{When the interactions are staggered, positive semi-definiteness still guarantees exactly ferromagnetic ground states at arbitrary interaction strengths, as discussed in Ref.~\cite{Su_2018}.}
The electron feels repulsion $n_A U_A $ and $n_B U_B$ from the respective sublattices; flat Hartree dispersion occurs when these are equal. Thus for any $\delta$ (which determines $n_A$ and $n_B$), we can find an appropriate ratio $U_A/U_B$ that completely extinguishes the Hartree dispersion, namely $U_A/U_B = {\sqrt{1+4/\delta^2}-1}$. 
We would then predict that by tuning the staggering $U_A/U_B$, we can change the onset location of the polaron formation.
We test this in Fig.~\ref{fig:3}(b) and find that indeed, the location where polaron first forms precisely tracks the value of $\delta$ where the Hartree dispersion is flat.
At this point, it is worth remarking that if one had assumed a naive kinetic magnetism mechanism for polaron formation in the MT chain, the expectation would have been that the interaction-induced dispersion would \textit{aid} in polaron formation.
Instead, we find the opposite.\footnote{In this way, our model differs from that of Ref.~\cite{Westerhout_2022}, where interactions within a flat band lead to effective kinetic magnetism.}

\textit{Binding Mechanism and Role of Band Mixing}---We now aim to understand some features of the mechanism which binds the electrons and the magnons in the MT chain and, moreover, what sets the scale of the threshold interaction strength at which the polaron first forms at each $\delta$.
To do so, we make the following preliminary observation: some form of band mixing must play a role in the formation of the polaron.
The very existence of a minimal interaction strength for the onset of polaron formation implies this. 
Indeed, if band mixing played no important role, the physics of the system would be governed by the physics of the flat band Hamiltonian whose \textit{only} energy scale is $U$; it is thus not possible for one to see a threshold in this setting.
This leaves open the question what precisely band mixing is doing.
We will now show that band mixing plays an essential role in the binding mechanism and, when $\delta < \delta_c$ it softens the energetics of the electron sufficiently to allow for polaron formation.

Regarding the binding mechanism of the electron and hole, one can imagine two scenarios.
First, there could be an attractive interaction that arises purely from the flat band, similar to the quantum Hall case~\cite{Khalaf_2022}, with band mixing merely assisting by modifying single particle energetics.
Alternatively, the attractive interaction itself arises due to hybridization with the dispersive band.
We now show that the latter scenario is borne out.

To demonstrate this, we evaluate the action of the flat band Hamiltonian of Eq.~\eqref{eq:flatbandformalism} on the states of electrons and magnons to consider a putative flat-band binding. 
Following Ref.~\cite{Khalaf_2022}, if $\xi_{n,p}$ and $\phi_{n,p}(k)$ denote the magnon energy and wave function (as a function of internal momentum $k$ and magnonic band index $n$) respectively, this Hamiltonian acts on a state with electron momentum $k_0+q$ and magnon momentum $q$ as:
\begin{multline}
    H |k_0; q, n \rangle = [\xi_{n,q} + \epsilon_H(k_0 + q)] |k_0; q, n \rangle \\
    + \frac{U}{2N} \sum_{q'}  \lambda^*_{q'}(k_0 + q) W^{nm}_{q,q'} |k_0; q + q', m \rangle,
    \label{Heph}
\end{multline}
where $W^{nm}_{q,q'}$ encodes the magnon's internal scattering needed for interaction.
In the End Matter we prove rigorously that for \textit{any} inversion-symmetric flat band ferromagnetic model with onsite repulsions in any dimension, a flat Hartree dispersion implies that this coefficient $W$ \textit{also} vanishes identically for all magnon modes. 

It is thus striking that spin polarons form precisely near the parameter value where the Hartree dispersion vanishes, i.e., where electrons and magnons should not interact at all as per the flat-band-projected Hamiltonian. This makes it clear that the binding mechanism originates not from the projected flat-band terms but from mixing with the dispersive band.
While the total occupancy of the dispersive band in the polaron state remains numerically small—as low as $0.05\%$ of the spin down electrons at the onset at $\delta_c$ \cite{SM}—this weak hybridization is nevertheless essential for binding.

Beyond generating attraction, we identify one additional role band mixing plays in facilitating polaron formation: It softens the energetic costs of the constituent electron and spin flip. 
As we explored earlier, a strong Hartree dispersion impedes binding; in particular, for $\delta < \delta_c$, the Hartree dip traps the electron near a single momentum. Meanwhile in the same parameter regime, the Schrieffer-Wolff (SW) dispersion correction (see supplemental material~\cite{SM} for a detailed discussion) tends to show a \textit{peak} at $\pi/2$ momentum. Thus we expect that at sufficiently large $U$, the SW peak [see orange curve in Fig.~\ref{fig:3}(c)] can overcome the Hartree dip [Fig.~\ref{fig:3}(a)].

Interestingly, this band-mixing-assisted softening picture allows us to \textit{predict} the interaction threshold for polaron formation at $\delta <\delta_c$.
To do so, we can crudely suppose that for sufficient $U$, the $\mathcal{O}(U^2/t)$ Schrieffer-Wolff peak can cancel out the $\mathcal{O}(U)$ Hartree dip. An estimate for this threshold is the $U$ value for which the peak height and Hartree dip depth are equal. The extracted $U$ are plotted in dotted line Figure~\ref{fig:numerical-evidence}(b) and show remarkable agreement with the actual polaron threshold both qualitatively and quantitatively at small $U$ above the threshold. While this threshold scale at $\delta_c$ is set by dispersive-band-mixed interactions beyond the present analysis, the main contribution to the threshold $U$ below $\delta_c$ appears to be explained by the softening of electronic dispersion. This heuristic does not work as well for $\delta>\delta_c$, for which Schrieffer-Wolff dispersion is much weaker and lacks the apparent cancellation in shape with Hartree dispersion.

\textbf{Conclusions and Outlook.} In this work, we discovered the formation of a spin polaron in a context outside of the conventional paradigms of polaron formation: kinetic magnetism and topological bands.
In particular, our polarons arise in the Mielke-Tasaki model, which has no kinetic energy nor relevant band topology.
More surprisingly, these spin polarons form when the interaction-induced kinetic energy is minimized (in contrast to expectations from kinetic magnetism) and bind due to hybridization with a dispersive band (outside the usual flat band paradigm).
These findings motivate several interesting future directions.

In particular, there are several ways that would be interesting to extend our understanding of polaron formation in the MT chain.
Namely, given the key role played by the dispersive band for binding, it would be valuable to understand the precise Hamiltonian structure of the attractive potential between electrons and spin-flips generated from band mixing.
Even more interesting would be to see if this dispersive band is \textit{necessary}---could the quantum geometry of a topologically trivial band alone lead to a binding mechanism for electrons and spin-flips?

Moreover, it could also be fruitful to generalize our investigations to other contexts.
In particular, one could explore polaron formation in  higher-dimensional generalizations of the Mielke-Tasaki chain~\cite{Mielke_1992}, which also host flat bands and exhibit exact ferromagnetic ground states.
Intriguingly, in potential generalizations of our study to the bilayer context, interlayer antiferromagnetism could lead to Cooper pairing of polarons between opposite layers and the emergence of superconductivity \cite{Khalaf_2021, Chatterjee_2022, wang2025intertwinedordersquantumcriticality, wang2025spinpolaronmediatedsuperconductivitydoped, Sahay_2024}.

\textbf{Acknowledgments}
We thank Debanjan Chowdhury, Margarita Davydova, Patrick Ledwith, Johannes Mitscherling, Lev Kendrick, Eslam Khalaf, Xueyang Song, and Xuepeng Wang. A.V. is supported by the Simons Collaboration on Ultra-Quantum Matter, which is a grant from the Simons Foundation (651440, A.V.). 
For part of this work, R.S. was supported from the U.S. Department of Energy, Office of Science, Office of Advanced Scientific Computing Research, Department of Energy Computational Science Graduate Fellowship under Award Number
DESC0022158. S.P. was supported by the National
Science Foundation grant NSF DMR-2220703.

\small
\bibliographystyle{unsrt}
\bibliography{references}

\section*{End matter}

\textbf{Magnons in the flat band
.} In the flat band formalism, in a sector of total conserved momentum, electron-hole pairs atop the spin-up ferromagnetic background are described by basis states
\[|k;p\rangle \equiv c_{k\down}^\dagger c_{k+p \uparrow}|\textrm{FM}\rangle\]
The band-projected Hamiltonian has a simple action on this state:
\[\left(\hat H_{\rm flat} - \epsilon_H(k)\right) |k;p\rangle = U \int_0^{2\pi} \frac{dq}{2\pi}\lambda^*_q(k)\lambda_{q}(k+p)|k+q;p\rangle   \]
Its bound eigenstates are magnon modes $\sum_k \phi_{n,p}(k)\ket{k;p}$ with energies $\xi_{n,p}$, where $n$ enumerates the magnon band(s).
If we define $M_\sigma$ to be amplitude of the component of the magnon state consisting of on-site spin flip on a $\sigma$ sublattice site, i.e. forming the $2\times 2$ diagonal matrix
\begin{equation}
M \equiv \int_0^{2\pi}\frac{dq}{2\pi} \ket{u_q}\phi_{n,p}(k)\bra{u_{q+p}},
\end{equation}
then it follows that:
\begin{equation}
phi_{n,p}(k) = \frac{U}{E_H(k) - \xi_{p,n}}\langle u_k| M |u_{k+p}\rangle
\end{equation}
Thus the magnon wave function as a function of $k$ has this highly specific form. While one can solve it exactly, here we just remark on the consequences for the case of flat Hartree dispersion.

From \cite{Khalaf_2022}, the electron magnon interaction term in the flat band contains the factor 
\begin{align}
    W^{nm}_{q,q'} = \sum_k \phi^*_{m,q + q'}(k) [&\lambda_{q'}(k) \phi_{n,q}(k + q')\\
    &- \phi_{n,q}(k) \lambda_{q'}(k + q) ]
    \label{Cnmqq2}
\end{align}
Using this functional form of the magnon wave function and adjusting overall normalization, 

\begin{equation}
\begin{split}
        W^{nm}_{q,q'} =\sum_k &\bra{u_{k+q+q'}} M^*_{m,q+q'} \ket{u_k} \bra{u_k} u_{k+q'}\rangle \bra{u_{k+q'}} M_{n,q} \ket{u_{k+q+q'}} \\- &\bra{u_{k+q+q'}} M^*_{m,q+q'} \ket{u_k} \bra{u_k} M_{n,q} \ket{u_{k+q}} \bra{u_{k+q}} u_{k+q+q'}\rangle
    \label{Cnmqq2}
\end{split}
\end{equation}
If the model has spatial/temporal inversion symmetry, the two terms cancel exactly upon a re-indexing $k\to -k-q-q'$.

Staggered interaction strengths $U_A/U_B \equiv v \neq 1$ can also be conveniently captured in the same flat band formalism by replacing the form factors with 
$$\tilde \lambda_q(k) \equiv \langle u_k|\begin{pmatrix}
    \sqrt{v} & \\ & 1
\end{pmatrix} |u_{k+q}\rangle;$$
the results mathematically are analogous.

\newpage

\newpage

\pagebreak
\onecolumngrid
\appendix

\section{Mielke-Tasaki Chain: Definition and Single-Particle Physics}

The Mielke Tasaki chain is an on-site repulsive Fermi-Hubbard model on a sawtooth chain, with sites denoted as $2x+\sigma$ where integer $x$ labels the unit cell position and we refer to $\sigma= 0,1$ as the $A$ and $B$ sublattices. The Hamiltonian of the Mielke Tasaki Chain is given by
\begin{equation}
     H = +t\sum_{x} \left( \psi^{\dagger}_{2x} + \psi^{\dagger}_{2x + 2} + \delta \psi^{\dagger}_{2x+1} \right) \left( \psi_{2x} + \psi_{2x+2} + \delta \psi_{2x+1} \right) + U\sum_j n_{j\uparrow}n_{j\downarrow}.
\end{equation}
Here $\delta>0$ is a dimensionless tuning parameter. Crucially, the hopping coefficient $t$ is positive (which we set to $t=1$). 
This Hamiltonian is symmetric under spin rotation $SU(2)_S$ and charge $U(1)_Q$. Famously, the Hamiltonian features an exact ferromagnetic ground state for \textit{any} positive $U$. 
A graphical depiction of the Hamiltonian is shown in Fig.~\ref{fig:MTschematic}: 
\begin{figure}[H]
    \centering 
    \includegraphics[width = 0.4 \textwidth]{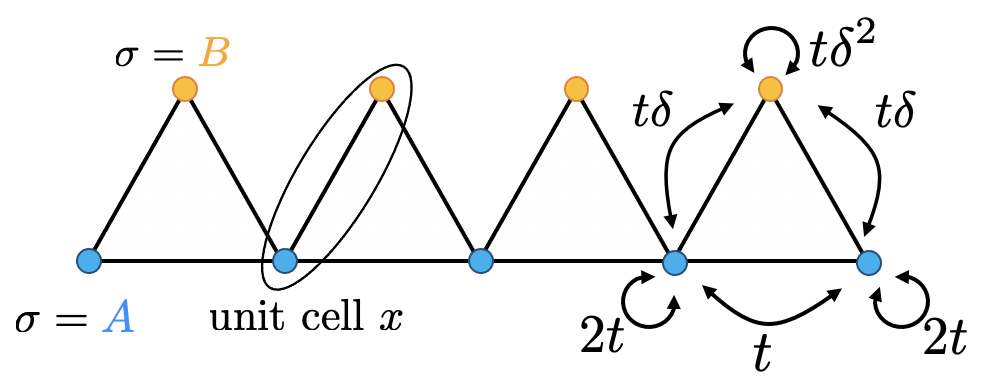}
    \caption{\textbf{The Mielke-Tasaki Chain.}}
    \label{fig:MTschematic}
\end{figure}


\subsection{Band Structure, Bloch Wavefunctions, and Wannier Functions }

We can diagonalize this Hamiltonian by going into the band basis.
\begin{equation}
    c_{k}^{\dagger} =  \psi^\dagger_{k} u_k  \qquad \psi_{k\sigma}^{\dagger} = \frac{1}{\sqrt{N}} \sum_{r} e^{i k \cdot (r + \sigma)}  \psi_{r, \sigma}^{\dagger} 
\end{equation}
where $[u_k]_{\sigma b}$ is a $2 \times 2$ matrix is the Bloch matrix in sublattice $\sigma$ and band $b$: 
\begin{equation}
    [u_k]_{\sigma b} = \frac{1}{\sqrt{\delta^2 + 4 \cos^2(ka)}}\begin{pmatrix} \delta & 2 \cos(ka) \\ -2 \cos(ka) & \delta \end{pmatrix}
\end{equation}
In this basis, the Hamiltonian takes the form: 
\begin{equation}
    H = \sum_{k} c_k^{\dagger} \varepsilon(k) c_k \qquad \varepsilon(k) = \begin{pmatrix}
        0 & \\ & 4 \cos^2(k) + \delta^2 
    \end{pmatrix}
\end{equation}
which means that the lowest band is perfectly flat and the single-particle gap is $\delta^2$. It is important to note that $k$ is defined in the Brillouin zone $[0,\pi)$. As a mathematical subtlety, the Bloch wave vector itself, as written above, is only single-valued for momenta in an \textit{extended} Brillouin zone $[0,2\pi)$, but the additional phases cancel out in physical quantities.

The localized Wannier orbitals on each unit cell $\varphi^\dagger_{2x} \equiv \sum_{j\in \mathbb{Z}} w(j) \psi^\dagger_{2x+j}$ can be found by Fourier transforming the Bloch wave functions.

\begin{equation}
        w(j) = \begin{cases} 
        \frac{1}{N}\sum_{k \in BZ} \frac{\delta}{\sqrt{\delta^2+4\cos^2 k}} e^{ikr}& \text{if } r\equiv 0 (\text{mod }2) \\
        -\frac{1}{N}\sum_{k \in BZ} \frac{e^{ik(r+1)}+e^{ik(r-1)}}{\sqrt{\delta^2+4\cos^2 k}} & \text{if }  r\equiv 1 (\text{mod }2)\\
        \end{cases}
\end{equation}

The Wannier functions asymptotically take the form $w(j) \propto e^{-\kappa |j|}/\sqrt{|j|} $ at long distances. Importantly for smaller $j$ there is a noticeably alternating weight between the $A$ and $B$ sublattices. The total supports of the Wannier function on the A sublattice and B sublattice is $\frac{\delta}{\sqrt{\delta^2+4}}$ and $1-\frac{\delta}{\sqrt{\delta^2+4}}$ respectively. These are precisely the $n_A$ and $n_B$ in the main text (there referring to the occupancy of the full ferromagnetic state).

\section{Additional Numerical Results for Polarons}

\subsection{Extended parameter regime}
\begin{figure}[h!]
    \centering
    \includegraphics[width=0.4\linewidth]{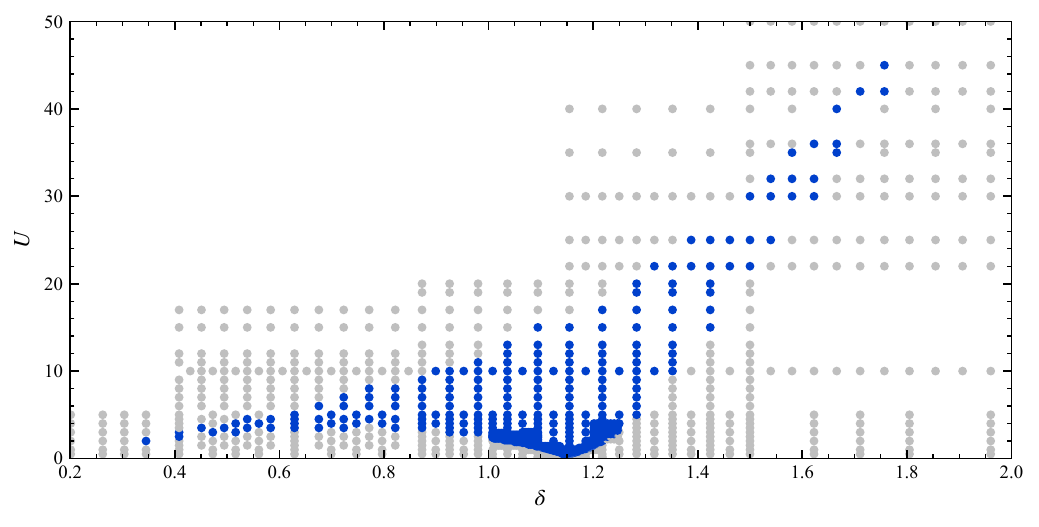}
    \caption{The presence of spin polarons was investigated in the parameter range $0.2\le \delta \le 2$ and $0 < U/t \le 50$. For $U/t\gtrsim 5$, the polaron binding energies decrease and the range of $\delta$ supporting the polaron also gets narrower while shifting to larger $\delta$. Polaron formation is observed to cease at $U/t \gtrsim 50$.}
    \label{fig:large_U}
\end{figure}

We have investigated spin polaron formation in a larger parameter range, including the limit of $U/t \gg 1$. The results are shown in Fig.\ref{fig:large_U}. Interestingly in the infinite $U$ limit there are no spin polarons; the highest $U/t$ for which they occur is $\sim 50$. Such large $U$ is a departure from explanations based around the physics of the flat band.

\subsection{Wave function analysis}

The main text discussed the continuous increase of binding energy with the onset of polaron formation at increasing $U$. This is also accompanied by a sudden discontinuous change in the wave funciton itself. Here we probe this by measuring the momentum-resolved occupancies

\[ n^{(b)}_{k,s} \equiv \langle c^\dagger_{k,s,b} c_{k,s,b}\]

here \(k\) is crystal momentum, \(s\in\{\uparrow,\downarrow\}\), and \(b\) labels the band.

In the \((N_\uparrow,N_\downarrow)=(N,1)\) sector and for \(U<U_c(\delta)\) (no polaron), the added electron is localized in momentum at the minimum of the Hartree-shifted single-particle dispersion. For \(\delta<\delta_c\) and \(\delta>\delta_c\) this yields a sharp peak of \(n^{(0)}_{k,\downarrow}\) at \(k=\pi/2\) and \(k=0\) respectively; at \(\delta=\delta_c\) the distribution remains narrow and strongly peaked around $k=0$ with only modest broadening. In the \((N_\uparrow,N_\downarrow)=(N-1,2)\) sector without binding, the ground state is that found by by acting with a spin-lowering operator on the electronic ground state, and the momentum-space occupancy mirrors that of the unbound electron.

\begin{figure}
    \centering
    \includegraphics[width=0.8\linewidth]{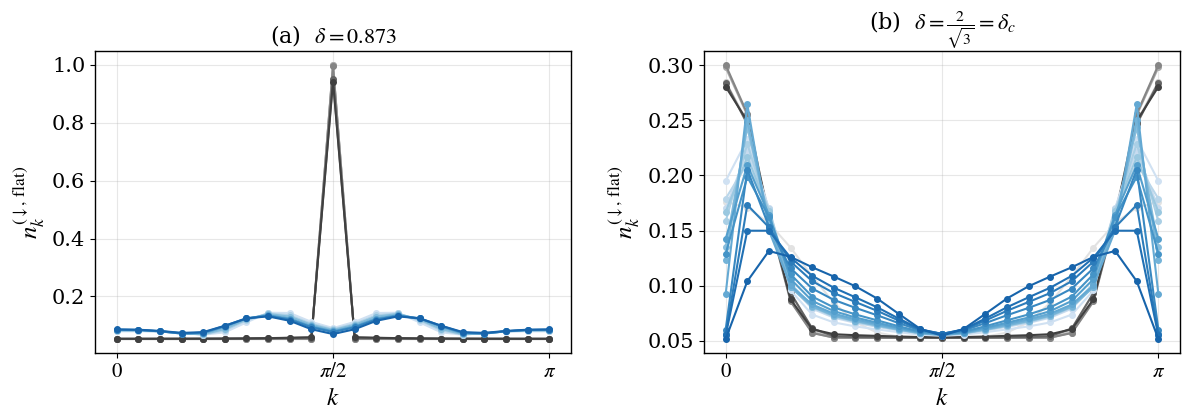}
    \caption{Momentum-space occupancy of the spin down electron in the one electron one spin flip sector. (a) For $\delta<\delta_c$: Without the polaron (grey) the momentum is peaked at the Hartree minimum. With the polaron (blue) there is an abrupt change and the full Brillouin zone is explored. (b) A similar but less pronounced change occurs at $\delta= \delta_c$.}
    \label{fig:n_k}
\end{figure}

Once a spin polaron forms for \(U>U_c(\delta)\), the momentum distribution reorganizes abruptly into a pronounced multi-\(k\) structure, reflecting an intricate linear combination of magnon and electron modes, shown in Figure \ref{fig:n_k}. Thus, whereas the binding energy turns on continuously at threshold, \(n^{(b)}_{k,s}\) exhibits a discontinuous reorganization, making momentum-space occupancy a sensitive diagnostic of polaron formation.

We also can compute the total occupancy of the dispersive band $\sum_{k,s} n_{k,s}^{\rm disp.}$ to quantify the deviation from the flat band regime. Generally, the onset of polaron formation coincides with a jump in dispersive band occupancy. The jump is small at $\delta_c$ but very noticeable away from $\delta_c$. Despite the smallness of the magnitude of $n_{\rm disp}$, we argue in the main text that this small mixing is essential for the binding glue between magnons and electrons.

\begin{figure}
    \centering
    \includegraphics[width=0.8\linewidth]{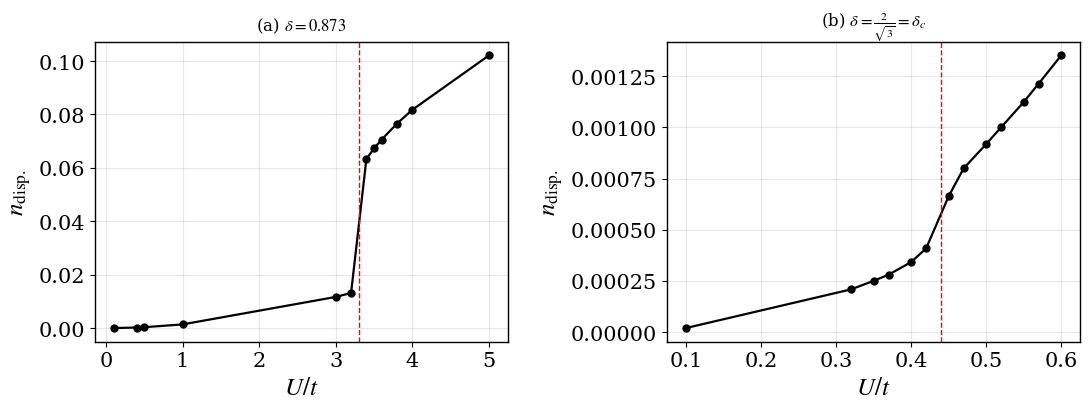}
    \caption{The total occupancy fo the dispersive band for the one electron one spin flip sector. Away from $\delta_c$, a sharp jump is observed at dispersive band occupancy at the onset of polaron formation. At $\delta_c$ this jump is less pronounced but also present.}
    \label{fig:dispocupancy}
\end{figure}

\section{Effects of Staggered Interaction Strength}

\
\begin{figure}[h!]
    \centering
    \begin{minipage}[t]{0.48\linewidth}
        \centering
        \includegraphics[width=0.8\linewidth]{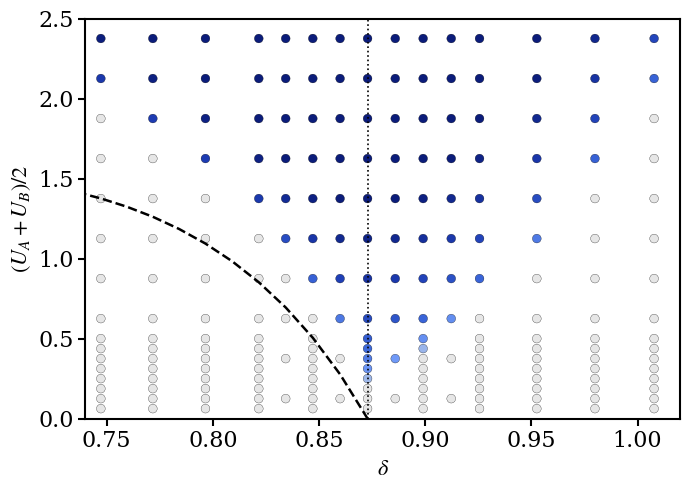}
        \caption{For $U_A/U_B = 1.5$ we plot the $\delta$ and $U$ values where nonzero polaron binding energy is observed (blue). The physical picture is similar to that of the uniform interaction case shown in the main text, except that the critical $\delta$ is shifted down. Using the Schrieffer Wolff heuristic we also estimate the threshold $U$ up to order of magnitude for $\delta$ below the critical value. }
        \label{fig:delta_vs_UAUB}
    \end{minipage}\hfill
    \begin{minipage}[t]{0.48\linewidth}
        \centering
        \includegraphics[width=0.8\linewidth]{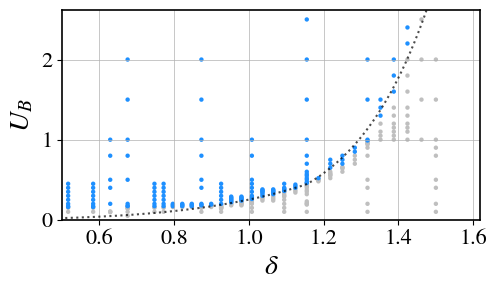}
        \caption{We simultaneously vary $\delta$ and $U_A/U_B$ tracking the regime of flat Hartree dispersion. In this regime, we plot the threshold interaction strength needed to see a spin polaron.}
        \label{fig:threshold_critical}
    \end{minipage}
\end{figure}

We can map out the parameter range of $U$ and $\delta$ where spin polarons form in the staggered itneraction case. Here we show the data for $U_A/U_B = 3/2$ in Fig.\ref{fig:delta_vs_UAUB}. 
As discussed in the main text, even in the regime of no Hartree dispersion, there is a small threshold interaction strength $U/t$. Here we show that threshold in \ref{fig:threshold_critical} as we tune through the critical $\delta$ regime by simultaneously adjusting the interaction strenght ratio $U_A/U_B \equiv v  = \sqrt{1+4/\delta^2}-1$.
The threshold is considerably lower than that with a Hartree dispersion. The threshold appears to follow the empirical relation $ U_B^{\rm min} \approx \left(\frac{1+v}{2+v}\right)^2/\min(v^3, v^5)$ for a range of parameter values, but a genuine understanding would require an investigation of the higher order interaction terms arising from band mixing.

\section{Schrieffer-Wolff corrections}

The flat-band-projected Hamiltonian is given by the following simple form:
\begin{equation}
    H_{\rm flat} + \frac{U}{2N}\sum_{q\in EBZ} \rho_{q\downarrow} \rho_{-q\uparrow} \qquad \rho_{q,s} = \lambda_q(k) c_{k,s}^\dagger c_{k+q,s}
\end{equation}

It represents a theory projected onto the low energy subspace of the non-interacting Hamiltonian, namely the flat band, with the interaction scale negligible compared to the band gap.

We are interested in the corrections to this picture upon introduction of a dispersive band $E_{\rm disp.}(k) = t\left(\delta^2+4\cos^2 k\right)$. This enters through treating the band mixing terms as perturbations to the non-interacting Hamiltonian.
This can be captured by the Schrieffer Wolff formalism.

\subsection*{Schrieffer Wolff Correction for Mielke Tasaki chain}

The general second–order correction schematically reads
\begin{equation}
H^{(2)}_{mn} = -\frac{2U^2}{T}\sum_p \frac{V^\dagger_{mp}V_{pn}}{E_p},
\end{equation}
where the original Hamiltonian term $V$ sends flat band states to dispersive band states. (While $E_p$ can be further replaced by $E_p-E_{m}$ we for simplicity take flat band energies to be zero to leading order). We decompose $V$ into three
contributions $V^{\downarrow} + V^{\uparrow}+V^{\uparrow \downarrow}$, shown here. For notational simplicity, we use $a_k=c_{\downarrow k}^{\rm flat}$, $b_k=c_{\uparrow k}^{\dagger \rm flat}$.
$A_k=c_{\downarrow k}^{\rm disp}$, and $B_k=c_{\uparrow k}^{\rm disp}$. We further have:
\begin{align}
-\,V^{\downarrow }
&= \frac{1}{2N}\sum_{k,k'\in \mathrm{BZ}}\sum_{q\in\mathrm{EBZ}}
\lambda^{10}_q(k)\,\lambda^{00}_{-q}(k')\;
A_k^\dagger\, a_{k+q}\, b_{k'}\, b_{k'-q}^\dagger \notag\\[2pt]
&= -\frac{1}{2N}\sum_{k,k'\in \mathrm{BZ}}\sum_{q\in\mathrm{EBZ}}
\lambda^{10}_q(k)\,\lambda^{00}_{-q}(k')\;
A_k^\dagger\, b_{k'-q}^\dagger b_{k'}\, a_{k+q}
+\tfrac{n_A-n_B}{2}\sum_{k\in \mathrm{BZ}} 
\lambda^{10}_\pi(k)\,A_k^\dagger a_k ,
\label{eq:Vdown}\\[6pt]
-\,V^{\uparrow }
&= \frac{1}{2N}\sum_{k,k'\in \mathrm{BZ}}\sum_{q\in\mathrm{EBZ}}
\lambda^{00}_q(k)\,\lambda^{10}_{-q}(k')\;
a_k^\dagger\, a_{k+q}\, B_{k'}^\dagger\, b_{k'-q}^\dagger ,
\label{eq:Vup}\\[6pt]
-\,V^{\uparrow \downarrow}
&= \frac{1}{2N}\sum_{k,k'\in \mathrm{BZ}}\sum_{q\in\mathrm{EBZ}}
\lambda^{10}_q(k)\,\lambda^{10}_{-q}(k')\;
A_k^\dagger\, a_{k+q}\, B_{k'}^\dagger\, b_{k'-q}^\dagger .
\label{eq:Vboth}
\end{align}

\noindent
Here the Bloch overlaps are
\begin{align}
\lambda^{00}_q(k) &= 
\frac{\delta^2 + 4\cos k\,\cos(k+q)}
{\sqrt{\delta^2+4\cos^2 k}\;\sqrt{\delta^2+4\cos^2(k+q)}},\\
\lambda^{10}_q(k) &=
\frac{2\delta\,[\cos k-\cos(k+q)]}
{\sqrt{\delta^2+4\cos^2 k}\;\sqrt{\delta^2+4\cos^2(k+q)}}\, .
\end{align}

For a single flat-band quasiparticle with momentum $p$, the Schrieffer-Wolff
second-order energy shift takes the form
\begin{equation}
E^{(2)}(p) \;=\; E^{(2)}_{\downarrow}(p)\;+\;E^{(2)}_{\uparrow}(p)\;+\;E^{(2)}_{\uparrow\downarrow}(p),
\end{equation}
corresponding respectively to virtual processes involving the (i) spin–down dispersive band only, (ii) spin–up dispersive band only, and (iii) both dispersive bands

\paragraph{(i) Spin–down dispersive band only.}
From the term in Eq.~\eqref{eq:Vdown}, the single–particle correction is diagonal and evaluates to
\begin{equation}
\boxed{~
E^{(2)}_{\downarrow}(p)
\;=\; -\,\frac{2U^2}{t} \,
\frac{4(n_A-n_B)^2\,\delta^2\,\cos^2 p}
{\Bigl(\delta^2+4\cos^2 p\Bigr)^3}
~}
\end{equation}

Notice that this vanishes identically at critical $\delta_c = 2/\sqrt{3}$ when $n_A= n_B = 1/2$.

\paragraph{(ii) Spin–up dispersive band only.}
From the term in Eq.~\eqref{eq:Vup}, the contribution can be written compactly as
\begin{equation}
\boxed{~
E^{(2)}_{\uparrow}(p)
\;=\; -\,\frac{2U^2}{t}\,
\frac{1}{(2N)^2}\!
\sum_{k\in[0,\pi)}\sum_{q\in[0,2\pi)}\!
\sum_{q'=q\;\text{or}\;q+\pi}
\frac{\lambda^{10}_{q}(k)\,\lambda^{10*}_{q'}(k)\,
\lambda^{00}_{-q'}(p)\,\lambda^{00*}_{-q}(p)}{E^{\rm disp}_k}
~}
\end{equation}

\paragraph{(iii) Both dispersive bands.}
From the term in Eq.~\eqref{eq:Vboth}, we obtain
\begin{equation}
\boxed{~
E^{(2)}_{\uparrow\downarrow}(p)
\;=\; -\,\frac{2U^2}{t}\,
\frac{1}{(2N)^2}\!
\sum_{k\in[0,\pi)}\sum_{q\in[0,2\pi)}\!
\sum_{q'\equiv q\;(\mathrm{mod}\;\pi)}
\frac{\lambda^{10}_{q}(k)\,\lambda^{10*}_{q'}(k)\,
\lambda^{01}_{-q'}(p)\,\lambda^{01*}_{-q}(p)}
{E^{\rm disp}_k + E^{\rm disp}_{p-q}}
~}
\end{equation}

Computing these integrals results in the figures of the main text. The bandwidth of the Schrieffer-Wolff dispersion falls sharply with increasing $\delta$.

The approach can be generalized for $U_A/U_B \neq 1$ by absorbing factors of $\sqrt{v} \equiv \sqrt{U_A/U_B}$ into the component of the form factors corresponding to the $A$ sublattice, i.e.

\begin{align}
\lambda^{00}_q(k) &= 
\frac{\sqrt{v}\delta^2 + 4\cos k\,\cos(k+q)}
{\sqrt{\delta^2+4\cos^2 k}\;\sqrt{\delta^2+4\cos^2(k+q)}},\\
\lambda^{10}_q(k) &=
\frac{2\delta\,[\sqrt{v}\cos k-\cos(k+q)]}
{\sqrt{\delta^2+4\cos^2 k}\;\sqrt{\delta^2+4\cos^2(k+q)}}\, .
\end{align}

These corrections are used to estimate the threshold $U$ for $\delta<\delta_c$ in the case where $U_A/U_B = 1.5$ as depicted in the dotted line of Figure~\ref{fig:delta_vs_UAUB}.

\end{document}